\documentclass[preprint,aps]{revtex4}

\usepackage{graphicx}
\usepackage{dcolumn}
\usepackage{bm}

\begin{document}


\title{Non-ideality of quantum operations with the electron spin of a $^{31}$P donor in a Si crystal due to interaction with a nuclear spin system.}

\author{S. Saikin}
\email{ssaykin@clarkson.edu}
 \altaffiliation[also at ]{Physics Department, Kazan State University, Kazan 420008, Russia}

\author{L. Fedichkin}%
\email{leonid@clarkson.edu}
\affiliation{%
Center for Quantum Device Technology, Clarkson University, Potsdam, New York 13699-5720, USA }%

\date{\today}

\begin{abstract}
We examine a $^{31}$P donor electron spin in a Si crystal to be used for the purposes of quantum computation. The interaction with an uncontrolled system of $^{29}$Si nuclear spins influences the electron spin dynamics appreciably. The hyperfine field at the $^{29}$Si nuclei positions is non-collinear with the external magnetic field. Quantum operations with the electron wave function, i.e. using magnetic field pulses or electrical gates, change the orientation of hyperfine field and disturb the nuclear spin system. This disturbance produces a deviation of the electron spin qubit from an ideal state, at a short time scale in comparison with the nuclear spin diffusion time. For ${\rm{H}}_{{\rm{ext}}}  \approx 9 $ T, the estimated error rate is comparable to the threshold value required by the quantum error correction algorithms. The rate is lower at higher external magnetic fields.
\end{abstract}

\pacs{Valid PACS appear here}
\maketitle



The interest in the evolution of a single electron spin interacting with a system of uncontrolled nuclear spins (or spin bath) arose from proposed models of spin qubits and qubit-qubit manipulations \cite{Loss, Kventsel, Kane, Awschalom, Jiang, Hu, Band, Glasser}. The electrons confined in a quantum dot or in a hydrogen state around a shallow donor impurity are influenced by many nuclear spins through the hyperfine dipole-dipole interaction. At low temperatures, where spin-phonon decoherence \cite{Mozyrsky} is ineffective, the dominant mechanism for the electron spin decoherence is via this interaction \cite{Khaetskii, Efros, Saykin, Sousa}. Additional error in a qubit state appears from non-ideal quantum gate operations \cite{XHu}. In this work we consider the effect of the nuclear spin bath on results of the quantum operations.

The approximation of short bath correlation times and weak system-bath interaction, which leads to a Markovian-like equation for relaxation and decoherence, is hardly applicable for quantum computation. In that case the evolution of qubit at short times is important \cite{Privman}. Different terms of system-bath interaction and the bath dynamics influence the evolution of a single qubit at different time scales. In particular, for the spin qubit, interacting with a spin bath, several time scales can be defined \cite{Stamp}.

The irreversible dynamics of the nuclear spin bath determines the longest time scale. At low temperatures this can be explained in terms of spin diffusion \cite{Slichter}. The characteristic diffusion time is determined by the flip-flop transitions ($ I^1_+ I^2_-+I^1_- I^2_+ $) for two neighbouring nuclear spins. These processes are much slower in comparison with the electron spin dynamics. For shorter time intervals, the electron spin sees a "frozen" system of nuclear spins. The diagonal component, $S_z I_z$, of the hyperfine interaction with a "frozen" spin system leads to the constant shift of the electron Larmor frequency. It is possible to eliminate this in quantum algorithms by various additional correction operations (in analogy with a  $\pi/2 - \pi$ pulse sequence), or by the preliminary measuring of the "nuclear magnetic field" at the localized electron. The single transitions with the electron spin flip ($ S_+ I_{\alpha}+S_- I_{\alpha} $, $ \alpha=+,-,z $) are suppressed by energy conservation, unless the external magnetic field is sufficiently low \cite{Khaetskii,Saykin}. However for nuclear spin flips, this suppression is not as strong, due to the small Zeeman splitting of nuclear levels. 

The effective magnetic field at each nuclear position is formed by the external and hyperfine fields. It is non-colinear to the external field, due to the hyperfine terms, $S_z I_+ +S_z I_-$. Quantum operations with the electron spin reorient this field, and cause the nuclear spin bath to oscillate near the nuclear Larmor frequency. These oscillations can be appreciably faster compared to nuclear spin diffusion. As a result, this produces a deviation of the electron spin qubit from an ideal state.  We study this effect in detail in the case of a $^{31}$P donor electron spin qubit system.
  
If, initially, at the time $t=0^-$  the qubit is uncoupled with the nuclear spin bath, and the bath is in an equilibrium state, $ \rho _{\sigma}(0^-) \rho _{bath}(0^-)$, after very fast quantum operation the system-bath state will be $ \rho _{\sigma}(0^+) \rho _{bath}(0^-)$. We assume here, that the qubit operation does not change the bath state significantly. The qubit density matrix at the time $t$ is
\begin{equation} \label{rho_t}
 \rho_{\sigma}(t)=\mathrm{Tr}_{bath}
 \left\{ e^{-\frac{i}{\hbar}Ht}
       \rho _{\sigma}(0^+) \rho _{bath}(0^-)
       e^{\frac{i}{\hbar}Ht} \right\}.
\end{equation}

At short times, in comparison with the nuclear spin diffusion, we neglect by the nuclear dipole-dipole interaction. The magnetic Hamiltonian of a single electron spin localized at a phosphorus donor in a $^{29}$Si$_x$ $^{28}$Si$_{1-x}$ crystal will be \cite{Hale}
\begin{equation} \label{Ham}
 H=g \mu_{\rm B} {\bf HS}+a_0{\bf SI}_0-\gamma _{\rm P} \hbar {\bf HI}_0 +\sum_k
 \left( {\bf SA}({\bf R}_k){\bf I}_k-\gamma _{\rm{Si}} \hbar {\bf HI}_k \right),
\end{equation}
where ${\bf I}_0$ is a donor nuclear spin and ${\bf I}_k$ are $^{29}$Si nuclear spins. The electron spin interaction with the phosphorus nucleus, the second term on the right-hand side of Eq.~(\ref{Ham}), is isotropic. The tensor, ${\bf A}({\bf R}_k)$, describes the hyperfine interaction of an electron spin with a nuclear spin of $^{29}$Si at a lattice position ${\bf R}_k$. It consists of isotropic and anisotropic parts as
\begin{equation} \label{Hypconst}
 A_{\alpha \beta}({\bf R}_k) = \frac{8}{3} \pi g_0 \gamma _{\rm{Si}} \hbar \mu _{\rm B} |\psi ({\bf R}_k)|^2 \delta _{\alpha \beta}+ D_{\alpha \beta}({\bf R}_k),
\end{equation}
where we use the free electron $g$-factor \cite{Paget}. Values of hyperfine constants have been measured by G. Feher \cite{Feher} and later by E. B. Hale and R. L. Mieher \cite{Hale} in the ENDOR experiments for different positions of nuclear spins. At some lattice positions the off-diagonal elements of tensor ${\bf A}({\bf R}_k)$, Eq. (\ref{Hypconst}), are comparable with the diagonal elements, Table \ref{tab:tt1}. Though the effective mass theory desribes symmetry of the hyperfine tensors incorrectly \cite{Feher}, the reasonable results have been obtained in the Bloch representations \cite{Ivey1, Ivey2} or in Wannier representations \cite{Onffroy} of a $^{31}$P donor electron wave function.

\begin{table}
\caption{\label{tab:tt1}Constants of the hyperfine interaction from the ENDOR experiments \cite{Ivey1} and numbers of equivalent positions, {\it n}, near the $^{31}$P donor.}
\begin{ruledtabular}
\begin{tabular}{lcccr}
 Shell & $|{\bf R}|$ (\AA) & $A_{zz}$ (KHz)&$A_{zx}$ (KHz) &$n$ \\
\hline
E (1 1 1)& 2.35 & 540 & 700 & 4 \\
B (4 4 0)& 7.68 & 4474 & -39.8 & 12 \\
H (4 4 -4)& 9.41 & 1378 & -50.6 & 12 \\
R (7 7 1)& 13.51 & 779 & 19.4 & 12 \\
\end{tabular}
\end{ruledtabular}
\end{table}

To calculate an evolution of the electron spin density matrix with the Hamiltonian, Eq. (\ref{Ham}), we use the approach derived by W.B. Mims \cite{Mims} for the modeling of electron spin echo envelope modulation functions (ESEEM). In a high magnetic field, we assume that the effect of the non-diagonal terms ($ S_+ I_{\alpha}+S_- I_{\alpha} $, $ \alpha=+,-,z $) in Eq. (\ref{Ham}) is small. These transitions are forbidden in the first order of perturbation and inversely proportional to the electron Zeeman level splitting in the second order. The truncated Hamiltonian can be written in the form
\begin{equation} \label{Htrunc}
 H_{\rm trunc}=g \mu_{\rm B} {\rm H} S_z+a_0 S_z I_z^0-\gamma _{\rm P} \hbar {\rm H} I_z^0+\sum_k
 \left( \sum_{\alpha} A_{z \alpha}({\bf R}_k)S_z I_{\alpha}^k-\gamma _{\rm{Si}} \hbar {\rm H}I_z^k \right).
\end{equation}
It is diagonal for the electron spin but non-diagonal for spins of $^{29}$Si nuclei. In the interaction representation with
\begin{equation} \label{Hint}
 H_0=g \mu_{\rm B} {\rm H} S_z+a_0 S_z I_z^0-\gamma _{\rm P} \hbar {\rm H} I_z^0,
\end{equation}
the Hamiltonian, Eq. (\ref{Htrunc}), can be diagonalized for different eigenvalues $m_s$ of $S_z$ using the rotation operator
\begin{equation} \label{Rot}
 S=\prod_k e^{i \varphi_k (m_s) I_y^k} e^{i \theta_k I_z^k},
\end{equation}
where $\tan \theta_k = A_{zy}({\bf R}_k)/A_{zx}({\bf R}_k)$, $A_{\bot}({\bf R}_k)=\sqrt{A_{zx}^2({\bf R}_k)+
A_{zy}^2({\bf R}_k)}$ and $\tan \varphi_k(m_s)=m_s A_{\bot}({\bf R}_k)/(A_{zz}({\bf R}_k)m_s-\gamma_{\rm Si} \hbar {\rm H})$.

We consider a $\pi/2$  magnetic pulse as the single qubit operation, $Op$, and assume that initially ($t=0^-$) the nuclear system is in equilibrium and unpolarized. The electron spin density matrix in the interaction picture, Eq. (\ref{Hint}), at time \textit{t} is,
\begin{eqnarray}
 \left( {\rho _\sigma  (t)} \right)_{m_s m'_s }  = \prod\limits_k {\left\{ {\cos ^2 \left( {\frac{{\varphi _k (m'_s ) -   \varphi _k (m_s )}}{2}} \right)\cos \left( {\frac{{\omega _k (m'_s )
  - \omega _k (m_s )}}{2}t} \right)} \right.} \nonumber \\
  +\sin ^2 \left( {\frac{{\varphi _k (m'_s ) - \varphi _k (m_s )}}{2}}
  \right)\left. {\cos \left( {\frac{{\omega _k (m'_s )   +\omega _k (m_s )}}{2}t}
  \right)} \right\}\left( {\rho _\sigma  (0^ +  )} \right)_{m_s m'_s },  \label{Ronon}
 \end{eqnarray}
where the effective oscillation frequency of the nuclear spin located at the lattice site \textit{k} with the electron spin projection $m_s$ is
\begin{equation} \label{Freq}
 \omega_k(m_s)=1/\hbar \sqrt{(A_{zz}({\bf R}_k)m_s-\gamma_{\rm Si} \hbar {\rm H})^2+
 \left(A_{\bot}({\bf R}_k)m_s \right)^2}.
\end{equation}

In a high magnetic field, where
\begin{equation} \label{Field}
 {\rm H} \gg |A_{zz}({\bf R}_k)/(2\gamma_{\rm Si} \hbar)|,
\end{equation}
this effect can be treated as a perturbation, and Eq. (\ref{Ronon}) is simplified to
\begin{eqnarray}
 \left( {\rho _\sigma  (t)} \right)_{m_s m'_s }  = \prod\limits_k {\Bigg\{ {\left( {1 - \frac{{\left( {A_ \bot  ({\bf{R}}_k )/2\hbar } \right)^2 }}{{(\gamma _{{\rm{Si}}} {\rm{H}})^2 }}} \right)^2 \cos \left( {\frac{{A_{zz} ({\bf{R}}_k )t}}{{2\hbar }}\left( {1 - \frac{{\left( {A_ \bot  ({\bf{R}}_k )/2\hbar } \right)^2 }}{{(\gamma _{{\rm{Si}}} {\rm{H}})^2 }}} \right)} \right)}
 }  \nonumber \\
+\frac{{\left( {A_ \bot  ({\bf{R}}_k )/2\hbar } \right)^2 }}{{(\gamma _{{\rm{Si}}} {\rm{H}})^2 }}\left. {\cos \left( {\gamma _{{\rm{Si}}} {\rm{H}} t\left( {1 - \frac{{\left( {A_ \bot  ({\bf{R}}_k )/2\hbar } \right)^2 }}{{(\gamma _{{\rm{Si}}} {\rm{H}})^2 }}} \right)} \right)} \right\}\left( {\rho _\sigma  (0^ +  )} \right)_{m_s m'_s }. \label{Rosim}
\end{eqnarray}

The reorientation of the effective magnetic field at each nuclear spin position changes the static component of the hyperfine field at the bounded electron, the first term in Eq. (\ref{Rosim}), and admixes oscillations near the $^{29}$Si nuclear Larmor frequency, the second term in Eq. (\ref{Rosim}). The amplitude and width of the admixed band are proportional to H$^{-2}$. This indicates that the effect can be reduced to any expected value by tuning of the external magnetic field. In the case of a polarized nuclear spin bath, the polarization $p_k$ for each nuclear spin, in analogy with a spin temperature, should be used. Therefore, Eq. (\ref{Ronon}) transforms to
\begin{eqnarray}
 \left( {\rho _\sigma  (t)} \right)_{m_s m'_s }  = \prod\limits_k \left( {\rho _\sigma  (0^ +  )} \right)_{m_s m'_s   }{\left\{ {\cos ^2 \left( {\frac{{\varphi _k (m'_s ) - \varphi _k (m_s )}}{2}}
 \right)\cos \left( {\frac{{\omega _k (m'_s ) - \omega _k (m_s )}}{2}t} \right)} \right.}  \nonumber \\
 +\sin ^2 \left( {\frac{{\varphi _k (m'_s ) - \varphi _k (m_s )}}{2}}
 \right)\cos \left( {\frac{{\omega _k (m'_s ) + \omega _k (m_s )}}{2}t} \right) \nonumber \\
 +ip_k \cos \left( {\frac{{\varphi _k (m'_s ) + \varphi _k (m_s )}}{2}} \right)
 \cos \left( {\frac{{\varphi _k (m'_s ) - \varphi _k (m_s )}}{2}} \right)\sin
 \left( {\frac{{\omega _k (m'_s ) - \omega _k (m_s )}}{2}t} \right) \nonumber \\
 \left. { - ip_k \sin \left( {\frac{{\varphi _k (m'_s ) + \varphi _k (m_s )}}{2}}
 \right)\sin \left( {\frac{{\varphi _k (m'_s ) - \varphi _k (m_s )}}{2}} \right)
 \sin \left( {\frac{{\omega _k (m'_s ) + \omega _k (m_s )}}{2}t} \right)}
 \right\}. \label{Ropol}
\end{eqnarray}

\begin{figure}
 \includegraphics{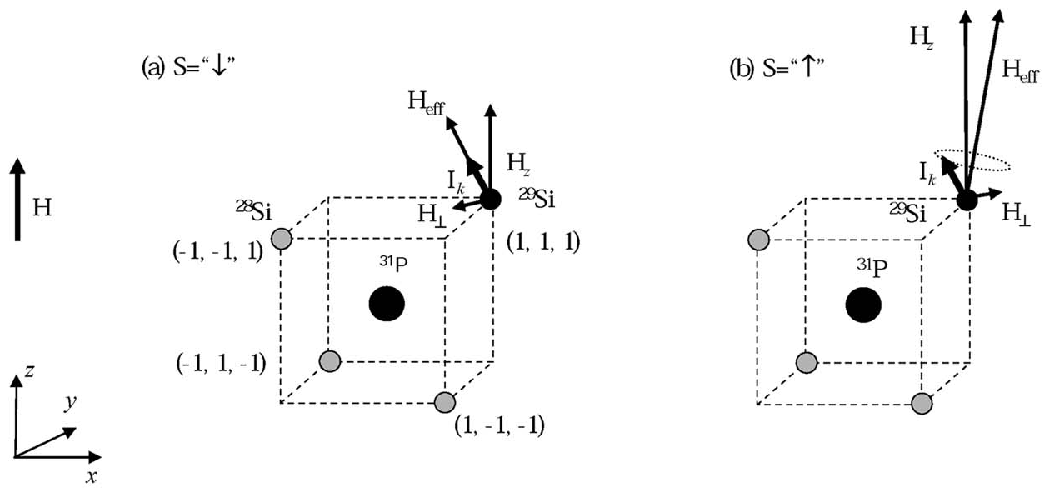}
\caption{\label{fig:ff1} The reorientation of the effective magnetic field  at the
lattice position (1,1,1) due to the electron spin flip. H$_z$ and H$_\bot$ are \textit{z} and in-plane components of the effective magnetic field H$_{\rm eff}$ at the $^{29}$Si position.}
\end{figure}

In a particular qubit implementation, the qubit density matrix
will depend on the position and polarization of each $^{29}$Si spin. We
estimate the value of this field reorientation effect for a
$^{31}$P electron spin interacting with a single $^{29}$Si nuclear
spin placed at one of the lattice sites from Table \ref{tab:tt1}
in an external magnetic field, H = 1 T.  The magnetic isotope
$^{29}$Si, which is located in one of the four, nearest to the
donor, lattice positions, experiences the largest reorientation of
the effective magnetic field due to the electron spin manipulation, Fig. \ref{fig:ff1}.
 If the nuclear spin is initially oriented parallel to $H_{\rm eff}$, Fig. \ref{fig:ff1}(a), 
then after the electron spin flips, it will rotate about new axis of quantization, Fig. \ref{fig:ff1}(b). To
quantify the influence of this rotation on the quantum bit evolution, we
calculated the norm of deviation of qubit state from ideal one \cite{Fedichkin}
\begin{equation} \label{Rodev}
||\Delta\rho_{\sigma}(t)||=||\rho_{\sigma}(t)- \rho_{\sigma}^{id}(t)||,
\end{equation}
where the deviation of the norm for single qubit is $||A||=\sqrt{A_{11}A_{22}+A_{12}A_{21}}$.
As the ideal state we used the spin polarized state, Eq. (\ref{Ropol}), without high frequency components.

\begin{figure}
\includegraphics{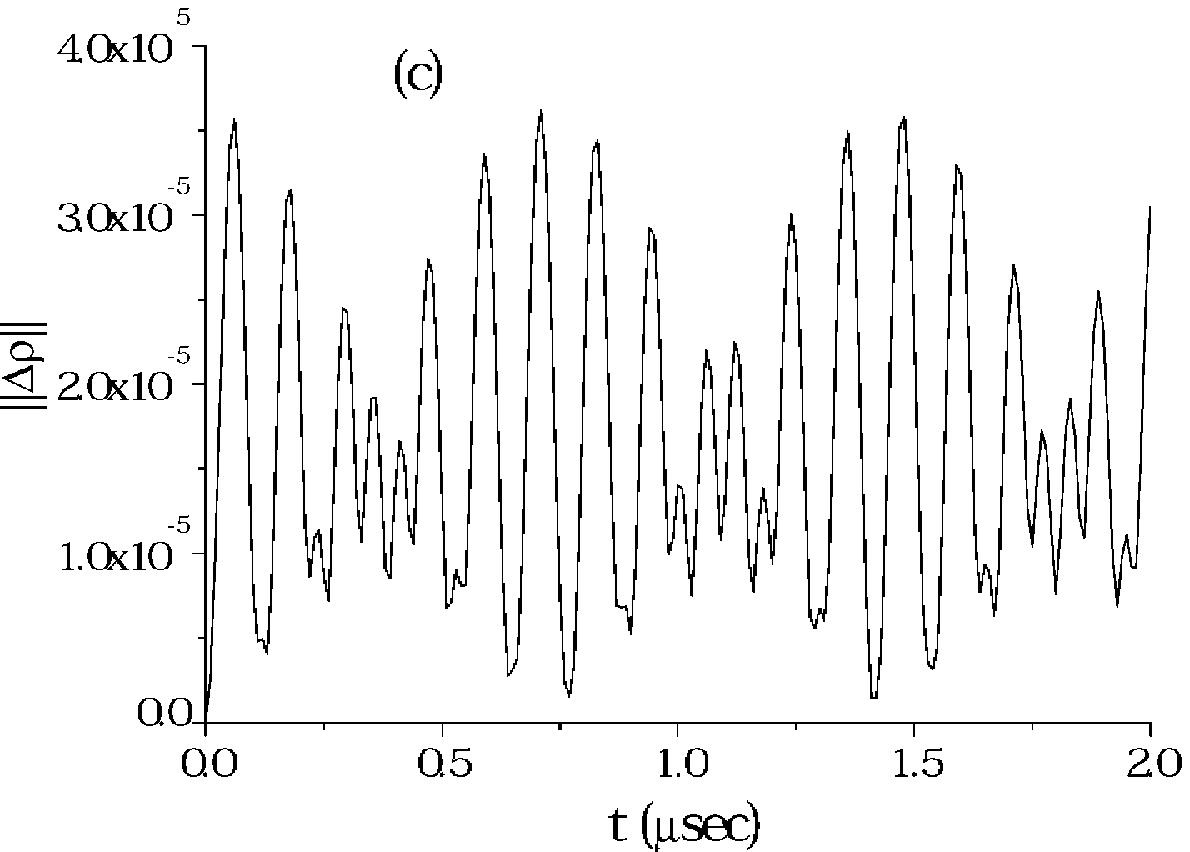}
\includegraphics{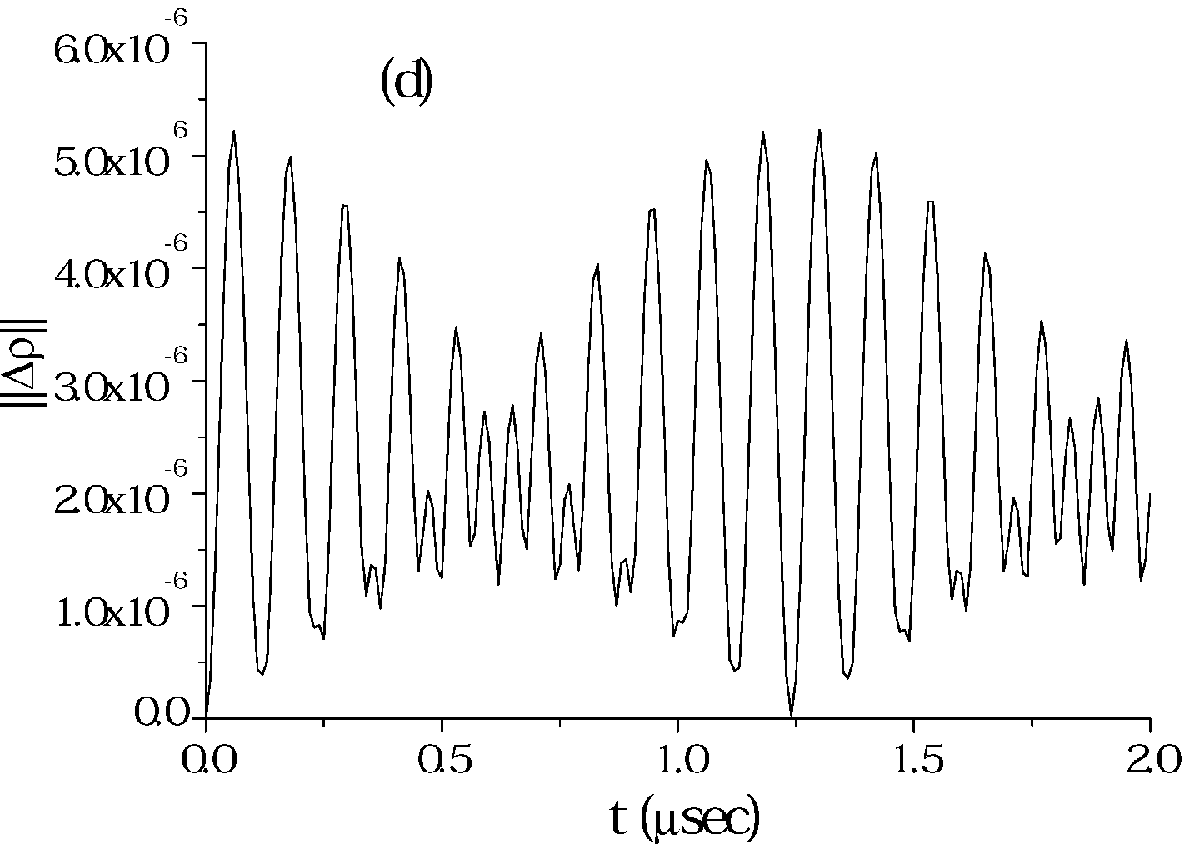}
\caption{\label{fig:ff2} The norm of the density matrix deviation
from the ideal state as a function of time for $^{29}$Si isotopes 
located in different lattice positions, see Table \ref{tab:tt1},
(a) position E, (b) position B, (c) postion H, (d) position R.
The external magnetic field H$=1$ T.}
\end{figure}

The simulated time evolution of the density matrix deviation,
after $Op=\pi /2$, is shown in Fig. \ref{fig:ff2}
for an initial nuclear spin polarization, $p_k=1$. For time scales on the
order of 100 ns, the amplitude of the density matrix deviation varies from 0 
to $7 \times 10^{ - 3} $, depending on the nuclear spin position.

To perform practically useful large scale quantum calculations, quantum error correction procedures should be used. The best known quantum error-correction schemes require the error rate to be less than one error per sequence of $ 10^3-10^5 $ quantum gates \cite{Preskill, Gott, DiV}. Our results indicate, for solid state nuclear spin quantum computers, the considered mechanism can be crucial, especially for qubits in which the $^{29}$Si isotopes are located at the nearest positions to the $^{31}$P.

For natural silicon crystal, n($^{29}$Si)$\approx 4.76\%$, the probability to have a $^{29}$Si in one of the four neighbors to the donor position is approximately $18\%$. Isotopic purification of a silicon crystal will reduce the probability to have a ``wrong'' qubit only. To suppress undesirable qubit dynamics, one should use high magnetic fields. It is possible to estimate the value of threshold magnetic field, using the $\left\| {\Delta \rho _{\max } } \right\| \sim {\rm{H}}^{ - 2} $ dependence, from $\left\| {\Delta \rho } \right\| \cdot Q = 1$ relation, where $Q \sim 10^4 $ is quality factor required for application of quantum error correction schemes. If we assume the effect of a whole nuclear magnetic bath is of the order of the maximal single nuclear spin effect, then the threshold value of the external magnetic field should be ${\rm{H}}_{{\rm{th}}}  \approx 9 $~T.

In the opposite case, to analyze this small effect, one should define experimental conditions in which it has considerable influence. To investigate the electron spin echo modulations, due to a nuclear spin at the E position (Table \ref{tab:tt1}), the magnetic field should be, at least, of the order of H$\approx0.1$ T (the electron Zeeman frequency $\nu_Z\approx2.8$ GHz). Although the ESEEM, due to interaction with $^{29}$Si nuclear spins, has been observed on the irradiation-induced defects in quartz at the electron Zeeman frequency $\nu_Z = 1.7$ GHz \cite{Kurshev}, as far as we know, there was no published works on spin echo modulations for the phosphorus donor electron spin.

In the case of GaAs or InGaAs quantum dots the concentration of nuclear magnetic isotopes is $100\%$. The electron spin interacts with a much larger set of a nuclear spins, and we expect more complicated dynamics of system. While there are no experimental values of hyperfine constants for GaAs and InGaAs quantum dots, it may be possible to estimate them theoretically. We suppose that, in analogy with silicon, calculations in the effective mass approximation would be incorrect and one should use, at least, the LCAO model.

In conclusion, we considered the effect of the non-diagonal terms in the dipole-dipole interaction of a $^{31}$P electron spin with a system of $^{29}$Si nuclei in a high magnetic field. Electron spin manipulations generate fast oscillations of the nuclear spin bath. Due to this effect the electron spin density matrix deviates from an ideal state. In an external magnetic field of $\rm H = 1$ T, the value of deviation is on the order of $7 \times 10^{ - 3} $. This is crucial for the implementation of practically useful quantum algorithms with electron spin qubit. The estimated threshold value of the magnetic field for a fault tolerant $^{31}$P electron spin qubit is ${\rm{H}}_{{\rm{th}}}  \approx 9$ T.

\begin{acknowledgments}
 We thank S. A. Lyon, J. Nesteroff and V. Privman for helpful discussions. This research was supported by the National  Security Agency and Advanced Research and Development Activity under Army Research Office contract DAAD-19-02-1-0035, and  by the National Science Foundation, grants DMR-0121146 and ECS-0102500.
\end{acknowledgments}

\begin{figure}
\includegraphics{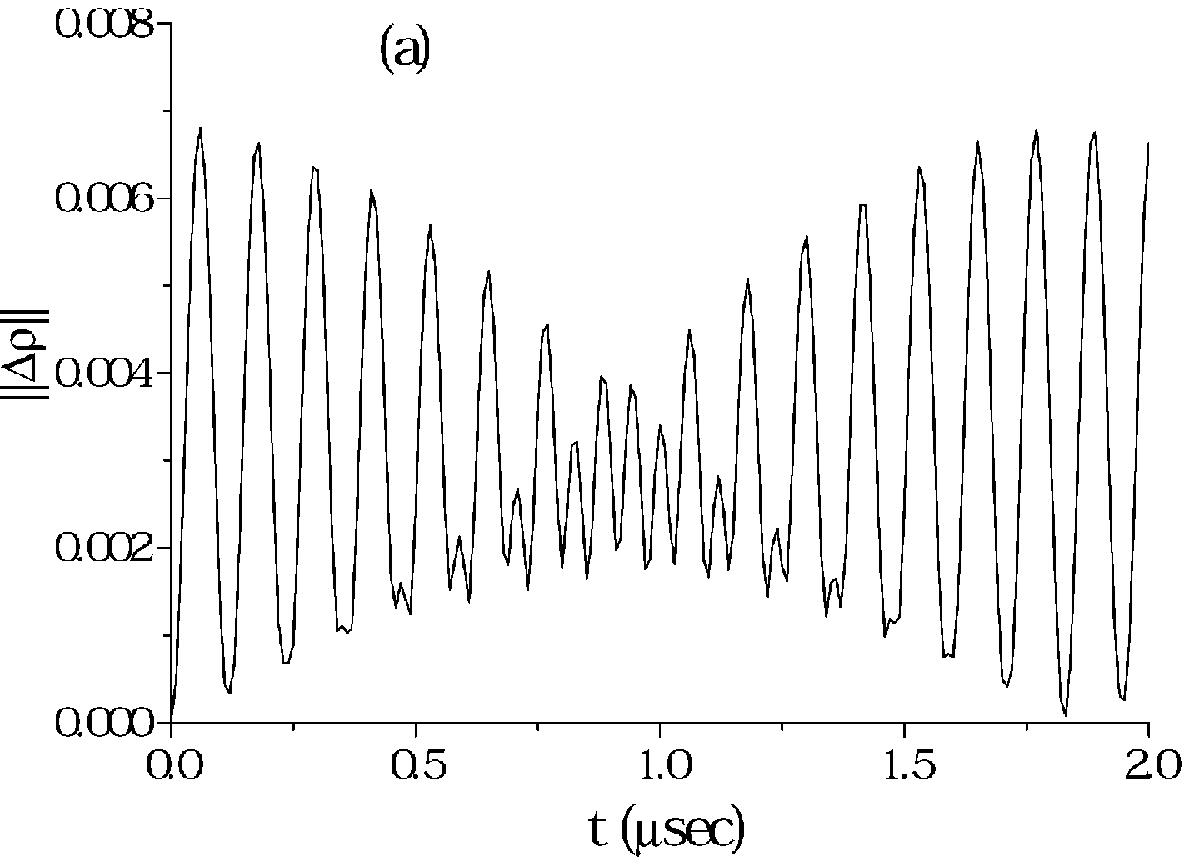}
\includegraphics{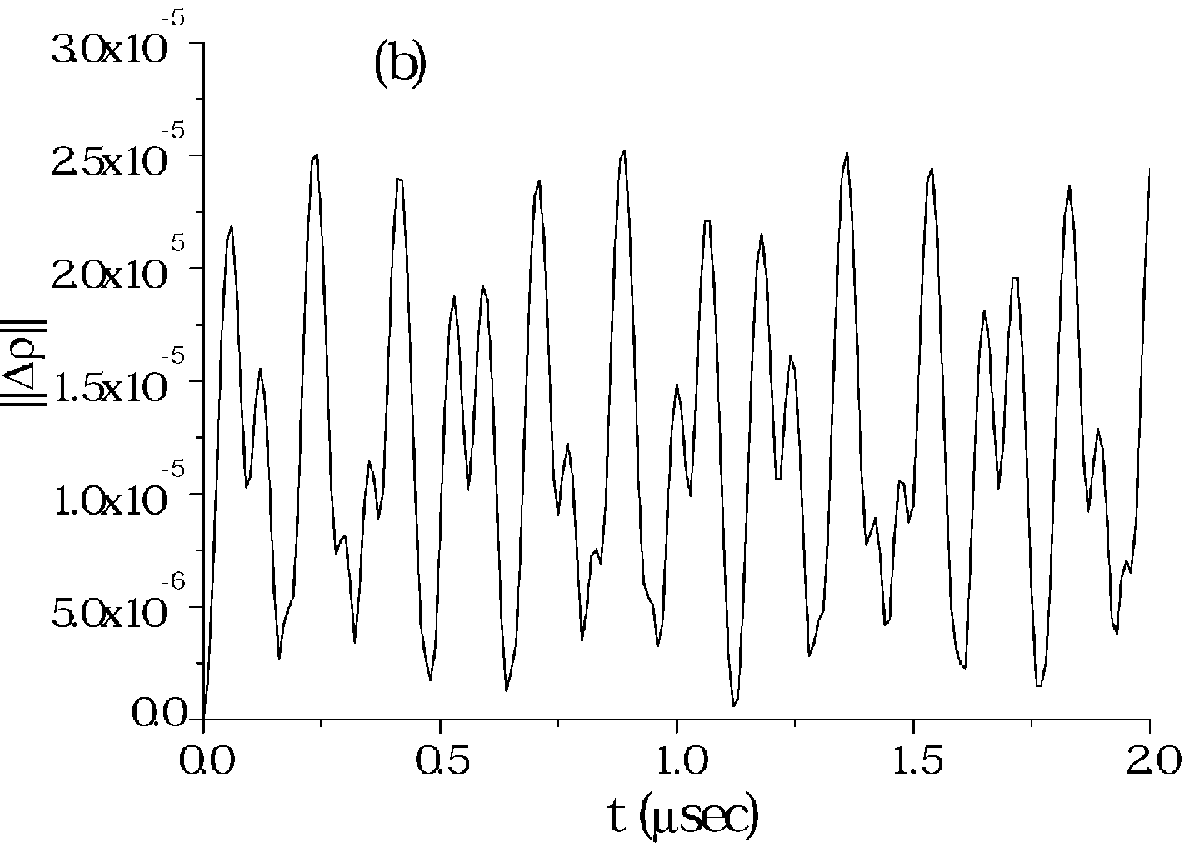}
\end{figure}

\end{document}